\documentstyle[aps,epsf,rotate,preprint]{revtex}

\begin{document}

\draft

\title{Multifractality in Human Heartbeat Dynamics}

\author{ Plamen Ch. Ivanov$^{1,2}$, Lu\'{\i}s A. Nunes Amaral$^{1,2}$, 
Ary L. Goldberger$^2$, \\
Shlomo Havlin$^3$, Michael G. Rosenblum$^4$,  
Zbigniew Struzik$^5$,  and H. Eugene Stanley$^1$}

\address{
$^1$Center for Polymer Studies and Department of Physics, 
        Boston University, Boston, MA 02215 \\
$^2$Harvard Medical School, Beth Israel Deaconess Medical Center, 
	Boston, MA 02215\\
$^3$Gonda Goldschmid Center and Department of Physics, 
	Bar-Ilan University, Ramat Gan, Israel\\
$^4$Department of Physics, Potsdam University, D-14415 Potsdam, Germany\\
$^5$Centre for Math. and Comp. Science, Kruislaan 413, 
    	NL-1098 SJ Amsterdam, The Netherlands}


\maketitle

\vspace*{0.5cm}

{\bf Recent evidence suggests that physiological signals under healthy
conditions may have a fractal temporal structure
\cite{Bassingthwaighte94}.  We investigate the possibility that time
series generated by certain physiological control systems may be
members of a special class of complex processes, termed multifractal,
which require a large number of exponents to characterize their
scaling properties \cite{Dewey,Stanley,Vicsek,Takayasu,Bunde}.  We
report on evidence for multifractality in a biological dynamical
system --- the healthy human heartbeat.  Further, we show that the
multifractal character and nonlinear properties of the healthy heart
rate are encoded in the Fourier phases.  We uncover a loss of
multifractality for a life-threatening condition, congestive heart
failure.
}

\vspace*{0.5cm}

\begin{figure}
\narrowtext
\vspace*{-1.0cm}
\centerline{\epsfysize=0.36\columnwidth{
{\epsfbox{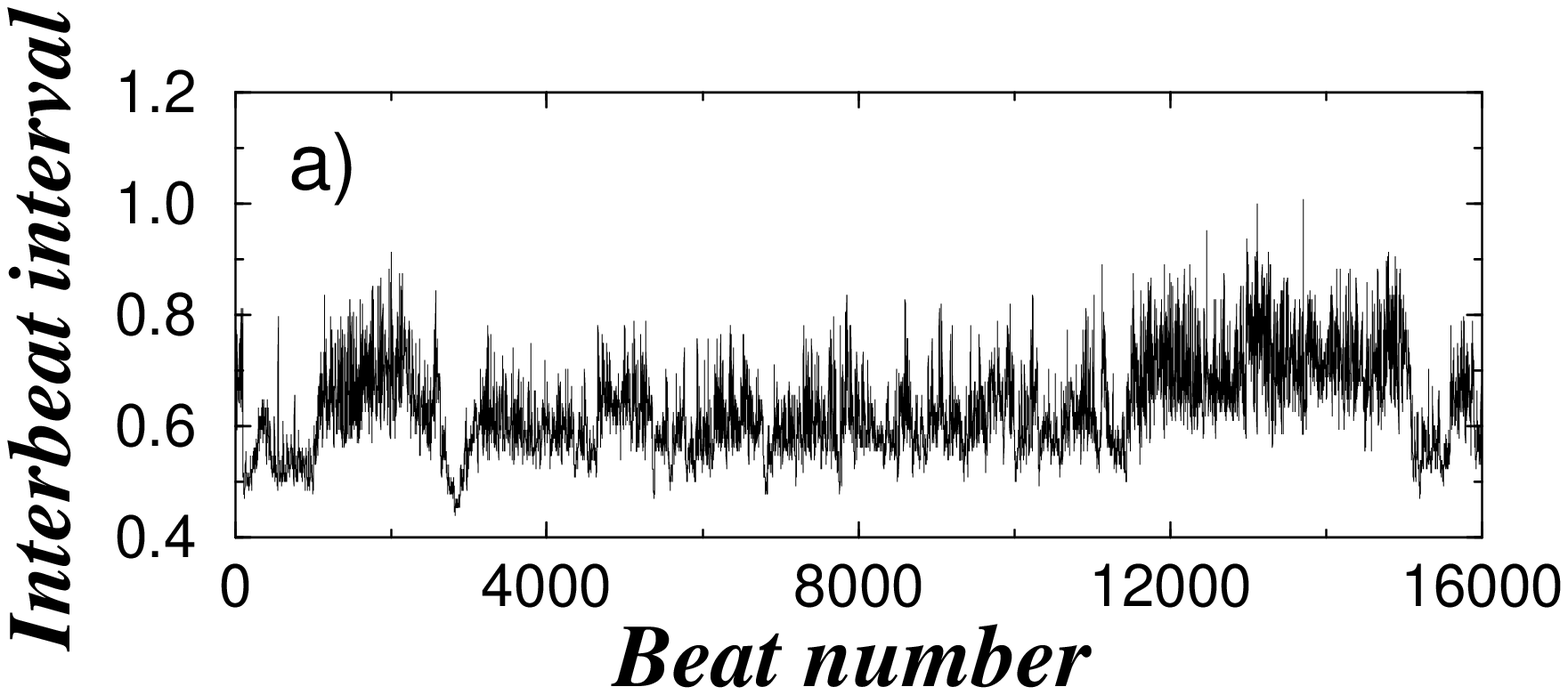}}}}
\vspace*{-0.5cm}
\centerline{\hspace*{2.0cm}\epsfysize=1.0\columnwidth{\epsfbox
{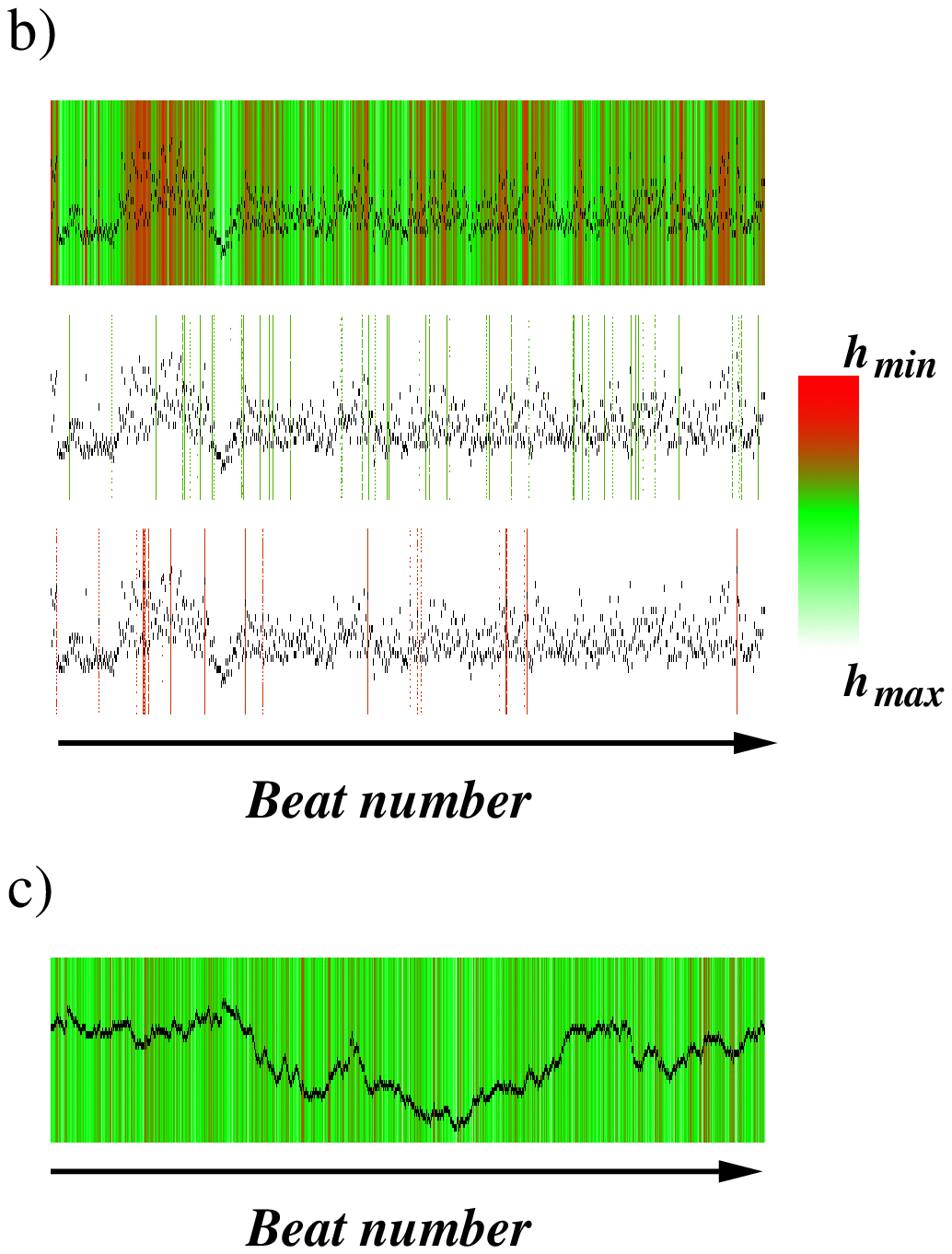}}}
\vspace*{0.5cm}
\vfill
\eject
\caption{ (a) Consecutive heartbeat intervals measured in seconds are
plotted vs beat number from approximately 3 hours record of a
representative healthy subject. The time series exhibits very
irregular and nonstationary behavior.  (b) The top panel displays in
color the local Hurst exponents calculated for the same 3 hours record
shown in (a). The heterogeneity of the healthy heartbeat is
represented by the broad range of local Hurst exponents $h$ (colors)
present and the complex temporal organization of the different
exponents. The middle and bottom panels illustrate the different
fractal structure of two subsets of the
time series characterized by different local Hurst exponents. The
value of the local Hurst exponent for each subset is represented with
a shade of green and red respectively. The two subsets display
different temporal structure which can be quantified by different
fractal dimension $D(h)$.  (c) The panel displays in color the local
Hurst exponents calculated for a {\it monofractal} signal ---
fractional Brownian motion with $H=0.6$.  The homogeneity of the
signal is represented by the nearly monochromatic appearance of the
signal which indicates that the local Hurst exponent $h$ is the same
throughout the signal and identical to the global Hurst exponent $H$.}
\label{f-rr}
\end{figure}

  Biomedical signals are generated by complex self-regulating systems
that process inputs with a broad range of characteristics
\cite{Shlesinger,Malik}.  Many physiological time series such as the
one shown in Fig.~\ref{f-rr}a, are extremely inhomogeneous and
nonstationary, fluctuating in an irregular and complex manner.  The
analysis of the fractal properties of such fluctuations has been
restricted to second order linear characteristics such as the power
spectrum and the two-point autocorrelation function.  These analyses
reveal that the fractal behavior of healthy, free-running
physiological systems is often characterized by $1/f$-like scaling of
the power spectra \cite{Akselrod81,Musha82,Hausdorff96,Peng95}.

  Monofractal signals are homogeneous in the sense that they have the
same scaling properties throughout the entire signal. Therefore
monofractal signals can be indexed by a single {\it global\/}
exponent---the Hurst exponent $H$ \cite{Hurst}.  On the other hand,
multifractal signals, can be decomposed into many subsets
characterized by different {\it local\/} Hurst exponents $h$, which
quantify the local singular behavior and thus relate to the local
scaling of the time series. Thus multifractal signals require many
exponents to fully characterize their scaling properties
\cite{Vicsek,Takayasu,Bunde}.

  The statistical properties of the different subsets characterized by
these different exponents $h$ can be quantified by the function
$D(h)$, where $D(h_o)$ is the fractal dimension of the subset of the
time series characterized by the local Hurst exponent $h_o$
\cite{Dewey,Vicsek,Takayasu,Bunde}.  Thus, the multifractal approach for
signals, a concept introduced in the context of multi-affine functions
\cite{VicsekTurbo91,Barabasi}, has the potential to describe a
wide class of signals that are more complex then those characterized
by a single fractal dimension (such as classical 1/f noise).

 We test whether a large number of exponents is required to
characterize heterogeneous heartbeat interval time series
[Fig.~\ref{f-rr}] by undertaking multifractal analysis.  The first
problem is to extract the local value of $h$. To this end we use
methods derived from wavelet theory \cite{Daubechies88}.  The
properties of the wavelet transform make wavelet methods attractive
for the analysis of complex nonstationary time series such as one
encounters in physiology \cite{Ivanov96}.  In particular, wavelets can remove
polynomial trends that could lead box-counting techniques to fail to
quantify the local scaling of the signal \cite{Muzy}.  Additionally,
the time-frequency localization properties of the wavelets makes them
particularly useful for the task of revealing the underlying hierarchy
that governs the temporal distribution of the local Hurst exponents
\cite{Muzy94}.  Hence, the wavelet transform enables a reliable
multifractal~analysis~\cite{Muzy,Muzy94}.  As the analyzing wavelet,
we use derivatives of the Gaussian function, which allows us to
estimate the singular behavior and the corresponding exponent $h$ at a
given location in the time series.  The higher the order $n$ of the
derivative, the higher the order of the polynomial trends removed and
the better the detection of the temporal structure of the local
scaling exponents in the signal.

  We evaluate the local exponent $h$ through the modulus of the maxima
values of the wavelet transform at each point in the time series.  We
then estimate the scaling of the partition function $Z_q(a)$, which is
defined as the sum of the $q^{th}$ powers of the local maxima of the
modulus of the wavelet transform coefficients at scale $a$
\cite{Muzy94}.  For small scales, we expect
%
\begin{equation}
Z_q(a) \sim a^{\tau(q)}\,.
\label{e-tau}
\end{equation}
%
For certain values of $q$, the exponents $\tau(q)$ have familiar
meanings.  In particular, $\tau(2)$ is related to the scaling exponent
of the Fourier power spectra, $S(f)\sim 1/f^{ \beta}$, as $\beta = 2 +
\tau(2)$.  For positive $q$, $Z_q(a)$ reflects the scaling of the
large fluctuations and strong singularities, while for negative $q$,
$Z_q(a)$ reflects the scaling of the small fluctuations and weak
singularities \cite{Vicsek,Takayasu}.  Thus, the scaling exponents
$\tau(q)$ can reveal different aspects of cardiac dynamics.

  Monofractal signals display a linear $\tau(q)$ spectrum, $\tau(q) =
qH - 1$, where $H$ is the global Hurst exponent.  For multifractal
signals, $\tau(q)$ is a nonlinear function: $\tau(q) = qh(q) - D(h)$,
where $h(q)\equiv d \tau(q)/d q$ is not constant. The fractal
dimension $D(h)$, introduced earlier, is related to $\tau(q)$ through
a Legendre transform,
%
\begin{equation}
D(h) = q h - \tau(q)\,.
\label{e-fa}
\end{equation}

\begin{figure}
\narrowtext
\centerline{
\epsfysize=0.52\columnwidth{\epsfbox
{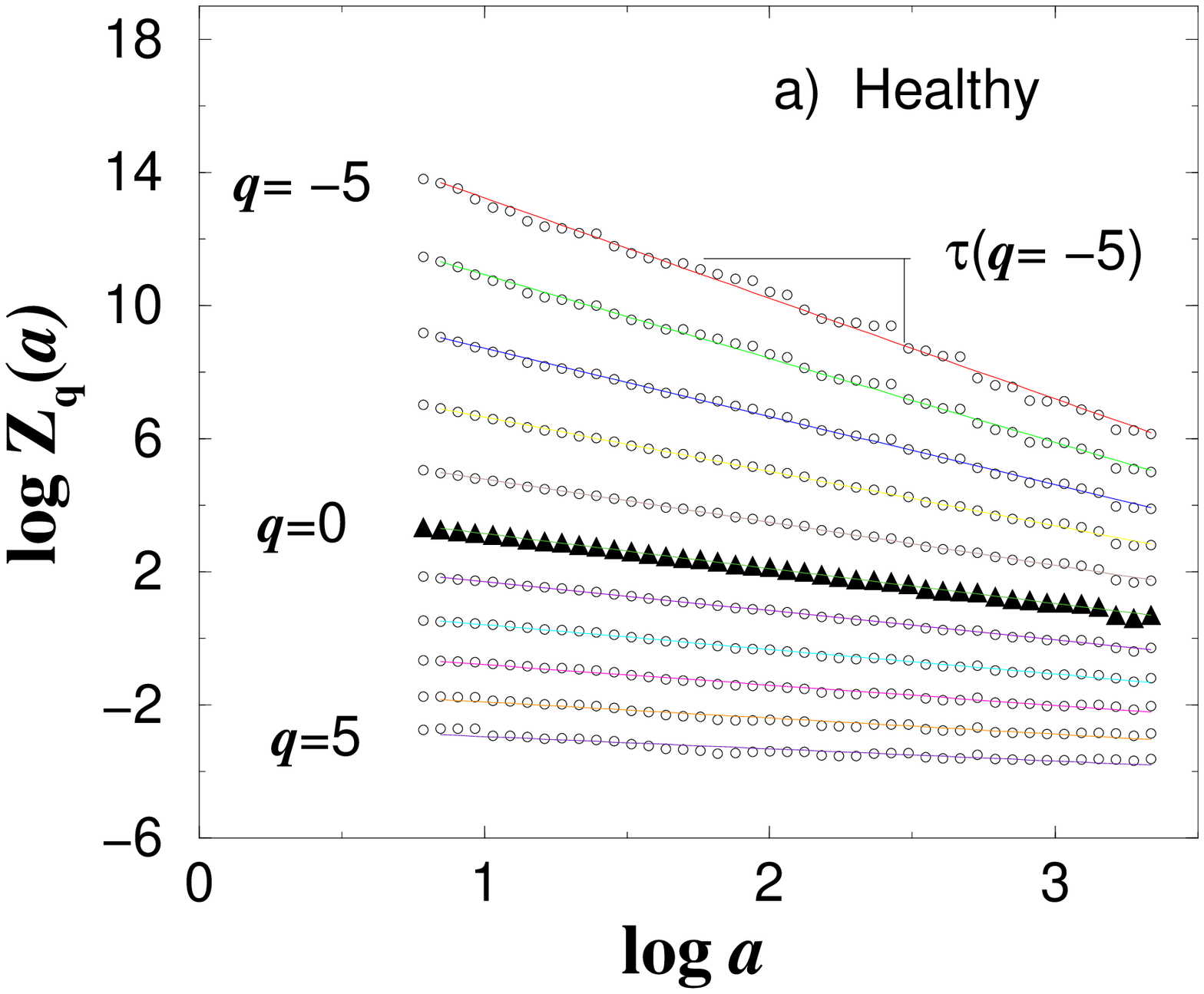}}
\hspace*{-1.0cm}
\epsfysize=0.52\columnwidth{\epsfbox
{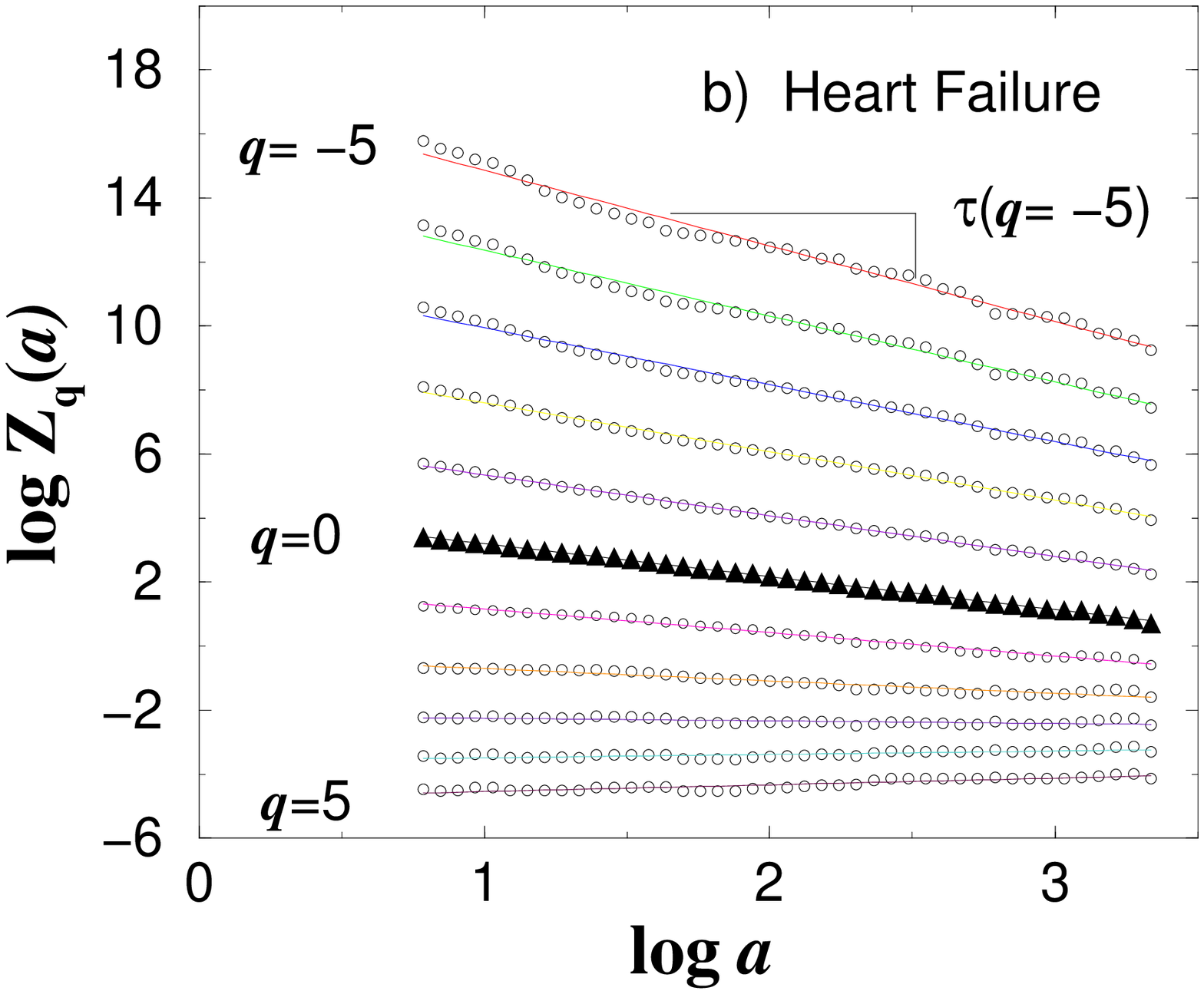}}}
\vspace*{0.5cm}
\centerline{
\epsfysize=0.52\columnwidth{\epsfbox
{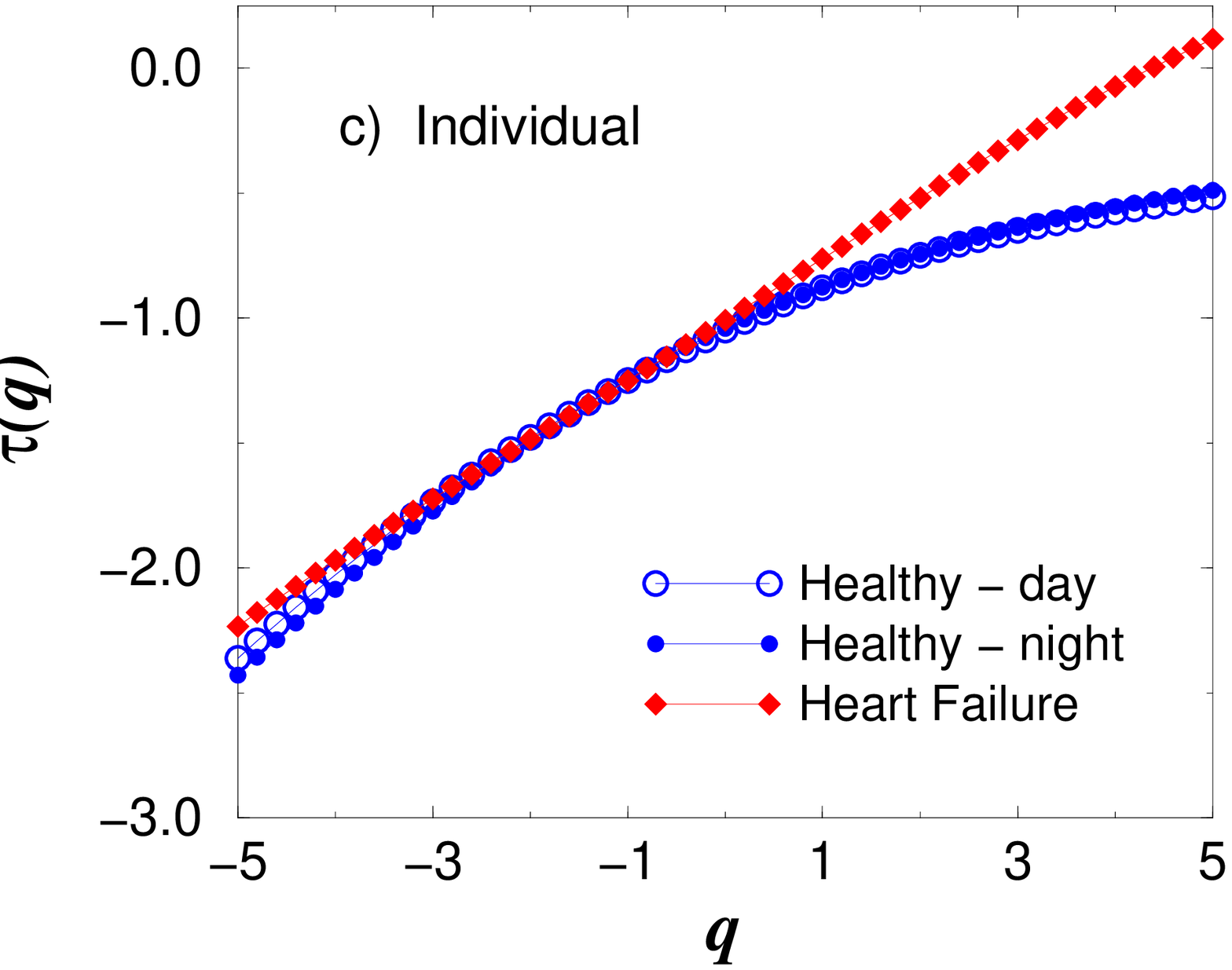}}}
\vspace*{0.5cm}
\caption{ Heartbeat time series contain densely packed, {\it
non-isolated\/} singularities which unavoidably affect each other in
the time-frequency decomposition.  Therefore, rather than evaluating
the distribution of the inherently unstable local singularity
exponents (as shown in color on Fig.~\protect\ref{f-rr}), we estimate
the scaling of an appropriately chosen global measure: the $q$ moments
of the probability distribution of the maxima of the wavelet transform
$Z_q(a)$ (as analyzing wavelet we use the 3rd derivative of the
Gaussian function).  Here we show the scaling of the partition
function $ Z_q(a)$ with scale $a$ obtained from daytime records
consisting of $\approx 25,000$ beats for (a) a healthy subject and (b)
a subject with congestive heart failure.  We calculate $\tau(q)$ for
moments $q = -5, 4, \ldots, 0, \ldots, 5$ and scales $a =
2\times1.15^i$, $i = 0, \ldots, 41$.  We display the calculated values
of $Z_q(a)$ for scales $a > 8$.  The top curve corresponds to $q=-5$,
the middle curve (shown heavy) to $q = 0$ and the bottom curve to
$q=5$. The exponents $\tau(q)$ are obtained from the slope of the
curves in the region {$16 < a < 700$}, thus eliminating the influence
of any residual small scale random noise due to ECG signal
pre-processing as well as extreme, large scale fluctuations of the
signal.  (c) Multifractal spectrum $\tau(q)$ for individual records.
A monofractal signal would correspond to a straight line for
$\tau(q)$, while for a multifractal signal $\tau(q)$ is nonlinear.
Note the clear differences between the curves for healthy and heart
failure records.  The constantly changing curvature of the curves for
the healthy records, suggests multifractality. In contrast, $\tau(q)$
is linear for the congestive heart failure subject, indicating 
monofractality.}
\label{f-scal}
\end{figure}

  We analyze both daytime (12:00 to 18:00) and nighttime (0:00 to
6:00) heartbeat time series records of healthy subjects, and the
daytime records of patients with congestive heart failure.  These data
were obtained by Holter monitoring \cite{Peng96}.  Our database
includes 18 healthy subjects (13 female and 5 male, with ages between
20 and 50, average 34.3 years), and 12 congestive heart failure
subjects (3 female and 9 male, with ages between 22 and 71, average
60.8 years) in sinus rhythm (see {\bf Methods} section for details on
data acquisition and preprocessing).  For all subjects, we find that
for a broad range of positive and negative $q$ the partition function
$Z_q(a)$ scales as a power law [Figs.~\ref{f-scal}a,b].

\begin{figure}
\narrowtext
\centerline{
\epsfysize=0.52\columnwidth{\epsfbox
{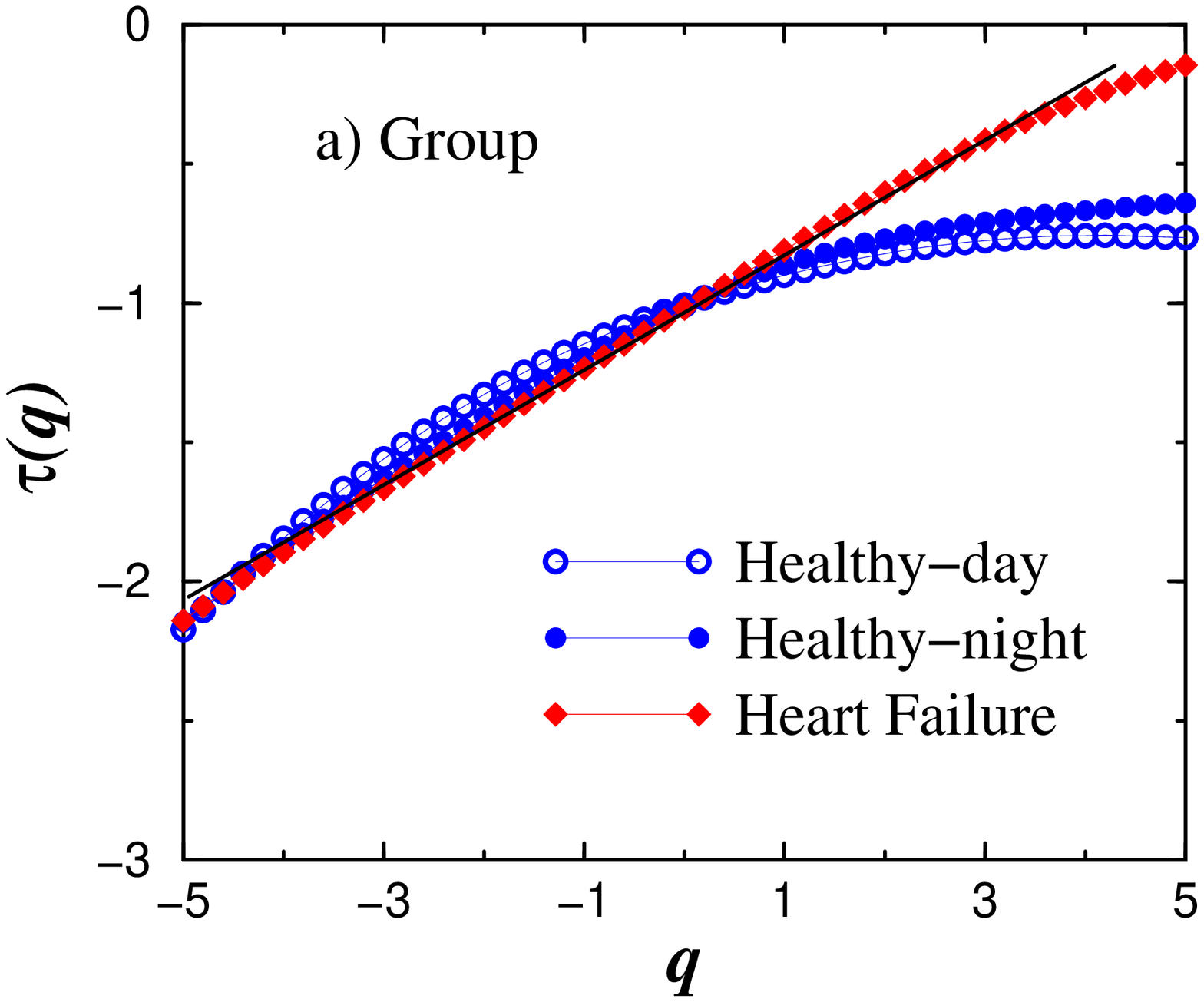}}
\hspace*{-1.0cm}
\epsfysize=0.52\columnwidth{\epsfbox
{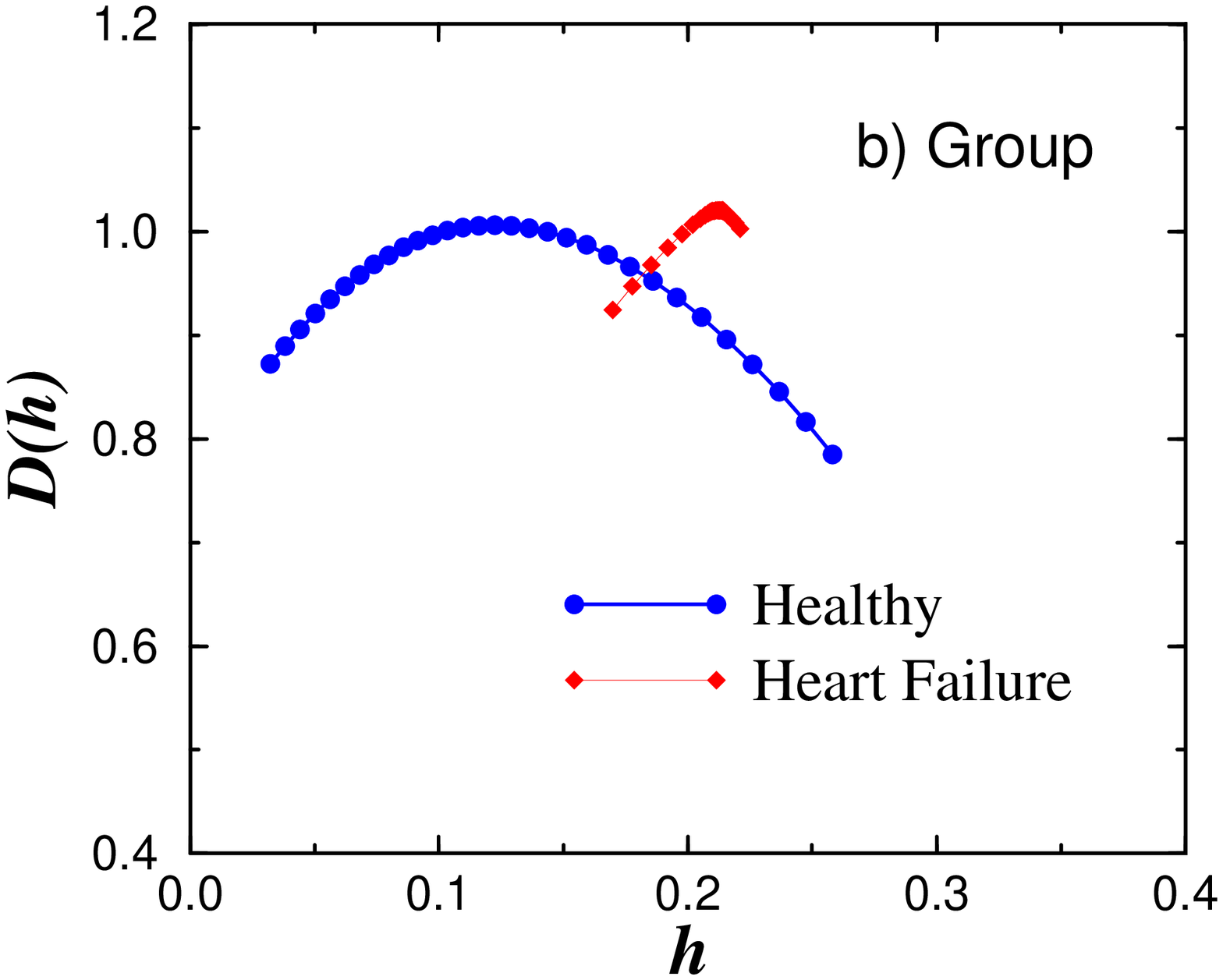}}}
\vspace*{1.0cm}
\caption{ (a) Multifractal spectrum $\tau(q)$ of the group averages
for daytime and nighttime records for 18 healthy subjects and for 12
patients with congestive heart failure.  The results show multifractal
behavior for the healthy group and distinct change in this behavior
for the heart failure group.  (b) Fractal dimensions $D(h)$ obtained
through a Legendre transform from the group averaged $\tau(q)$ spectra
of (a).  The shape of $D(h)$ for the individual records and for the
group average is broad, indicating multifractal behavior.  On the
other hand, $D(h)$ for the heart failure group is very narrow,
indicating monofractality.  The different form of $D(h)$ for the heart
failure group may reflect perturbation of the cardiac neuroautonomic
control mechanisms associated with this pathology.}  
\label{f-results}
\end{figure}

  For all healthy subjects, we find that $\tau(q)$ is a nonlinear
function [Fig.~\ref{f-scal}c and Fig.~\ref{f-results}a], 
which indicates that the heart rate
of healthy humans is a multifractal signal.  Figure~\ref{f-results}b
shows that for healthy subjects, $D(h)$ has nonzero values for a broad
range of local Hurst exponents $h$.  The multifractality of healthy
heartbeat dynamics cannot be explained by activity, as we analyze data
from subjects during nocturnal hours.  Furthermore, this multifractal
behavior cannot be attributed to sleep-stage transitions, as we find
multifractal features during daytime hours as well.  The range of
scaling exponents --- {$0 < h < 0.3$} --- with nonzero fractal dimension
$D(h)$, suggests that the fluctuations in the healthy heartbeat
dynamics exhibit anti-correlated behavior ($h = 1/2$ corresponds to
uncorrelated behavior while $h > 1/2$ corresponds to correlated
behavior).

  In contrast, we find that heart rate data from subjects with a
pathological condition --- congestive heart failure --- show a clear
loss of multifractality [Figs.~\ref{f-results}a,b].  For the heart
failure subjects, $\tau(q)$ is close to linear and $D(h)$ is non-zero
only over a very narrow range of exponents $h$ indicating
monofractal behaviour [Fig.~\ref{f-results}].

  Our results show that, for healthy subjects, local Hurst exponents
in the range {$0.07 < h < 0.17$} are associated with fractal
dimensions close to one.  This means that the subsets characterized by
these local exponents are statistically dominant.  On the other hand,
for the heart failure subjects, we find that the statistically
dominant exponents are confined to a narrow range of local Hurst
exponents centered at {$h \approx 0.22$}.  These results suggest that for heart
failure the fluctuations are less anti-correlated than for healthy
dynamics since the dominant scaling exponents $h$ are closer to $1/2$.

\begin{figure}
\narrowtext
\centerline{
\epsfysize=0.62\columnwidth{\epsfbox
{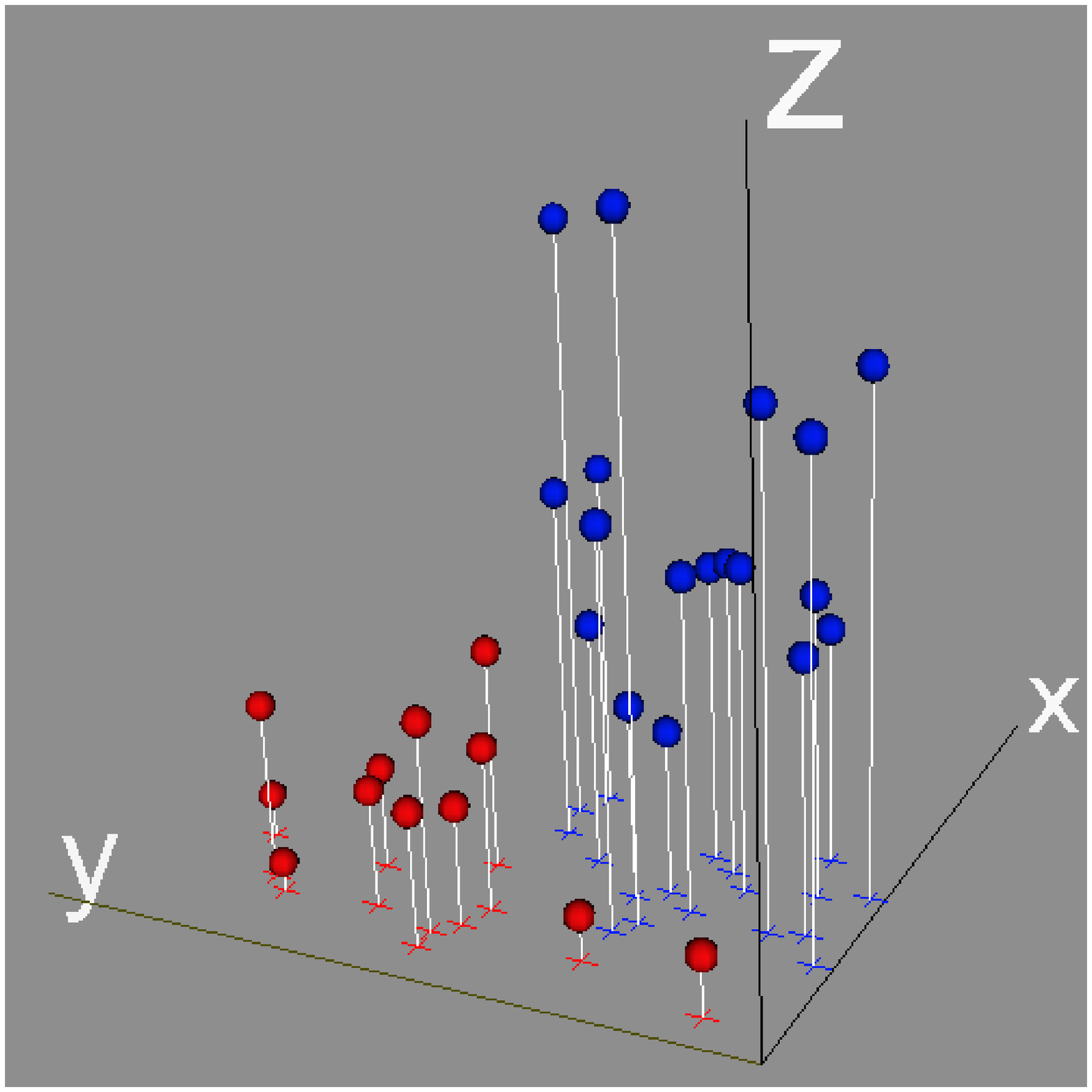}}}
\vspace*{1.5cm}
\centerline{
\epsfysize=0.52\columnwidth{\epsfbox
{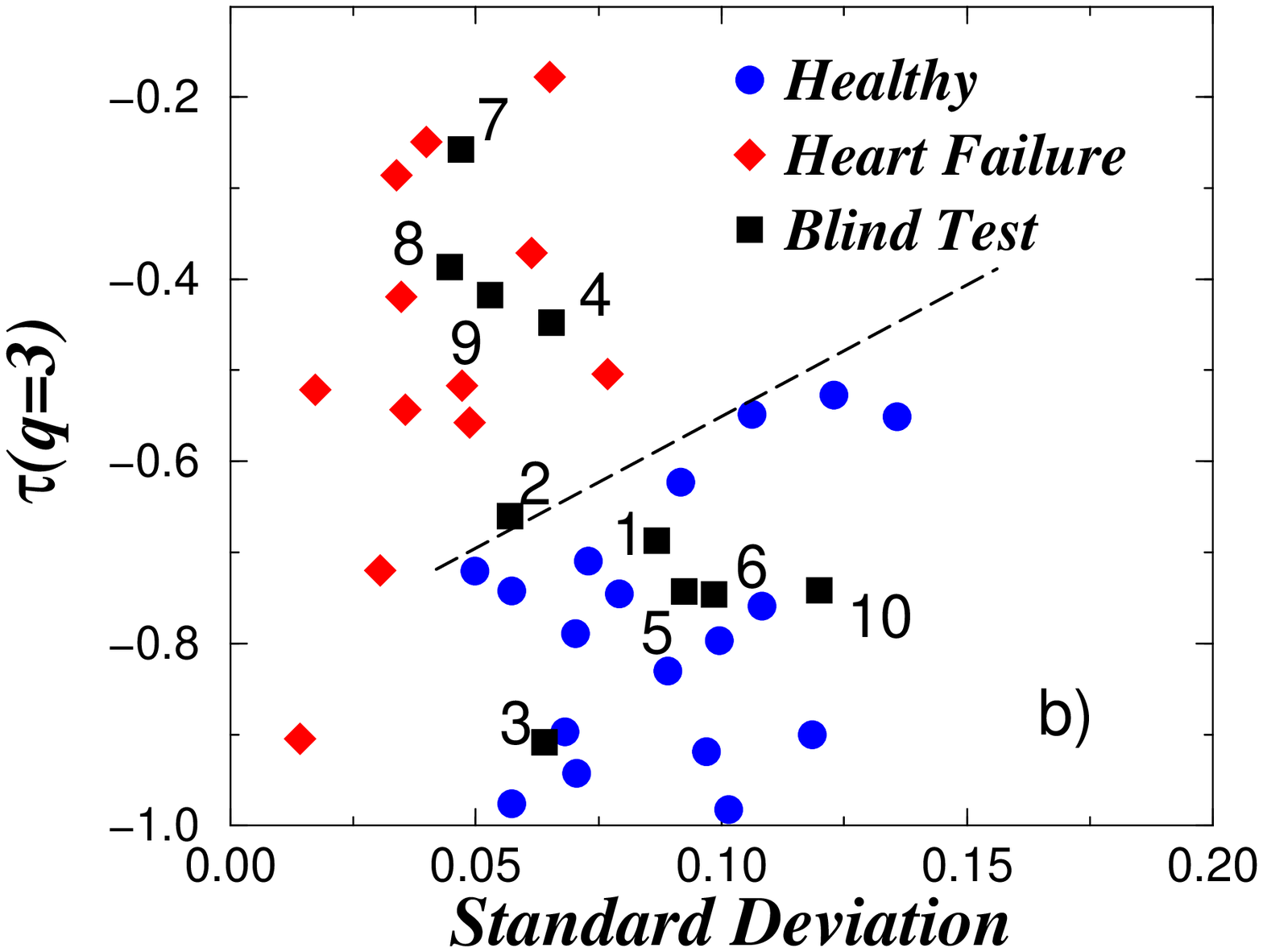}}}
\vspace*{1.0cm}
\caption{ (a) Discrimination
method based on the multifractal formalism. Each subject in the
database is characterized by three quantities. The first quantity
(z-axis) is the degree of multifractality, which is the difference
between the maximum and minimum values of local Hurst exponent $h$ for
each individual.  Note that the degree of multifractality takes value
zero for a monofractal.  The second quantity (y-axis) is the exponent
value $\tau(q=3)$ characterizing the scaling of the third moment
$Z_3(a)$. The third quantity (x-axis) is the standard deviation of the
interbeat intervals. The healthy subjects are represented by blue
spheres and the heart failure subjects by red spheres.  The figure
shows that the multifractal approach robustly discriminates the
healthy from heart failure subjects. (b) Discrimination method based
on multifractal formalism for the ``blind'' datasets.  The y-axis is
the exponent value $\tau(q=3)$, and x-axis is the standard deviation
of the time serie of interbeat intervals. Marked in black, we show the
results for the datasets of the blind test. We note that two
of the heart failure subjects (closest to the origin )
are ambiguous; however, they are clearly identified as unhealthy in Fig.~4a
because their degree of multifractality (z-axis) is close to zero
(these two subjects are the red spheres in Fig.~4a closest to the
origin). This case demonstrates our point that a single exponent is
not sufficient to describe the complexity of heartbeat time series.}
\label{f-discr}
\end{figure}
\begin{figure}
\narrowtext
\centerline{
\epsfysize=0.52\columnwidth{\epsfbox
{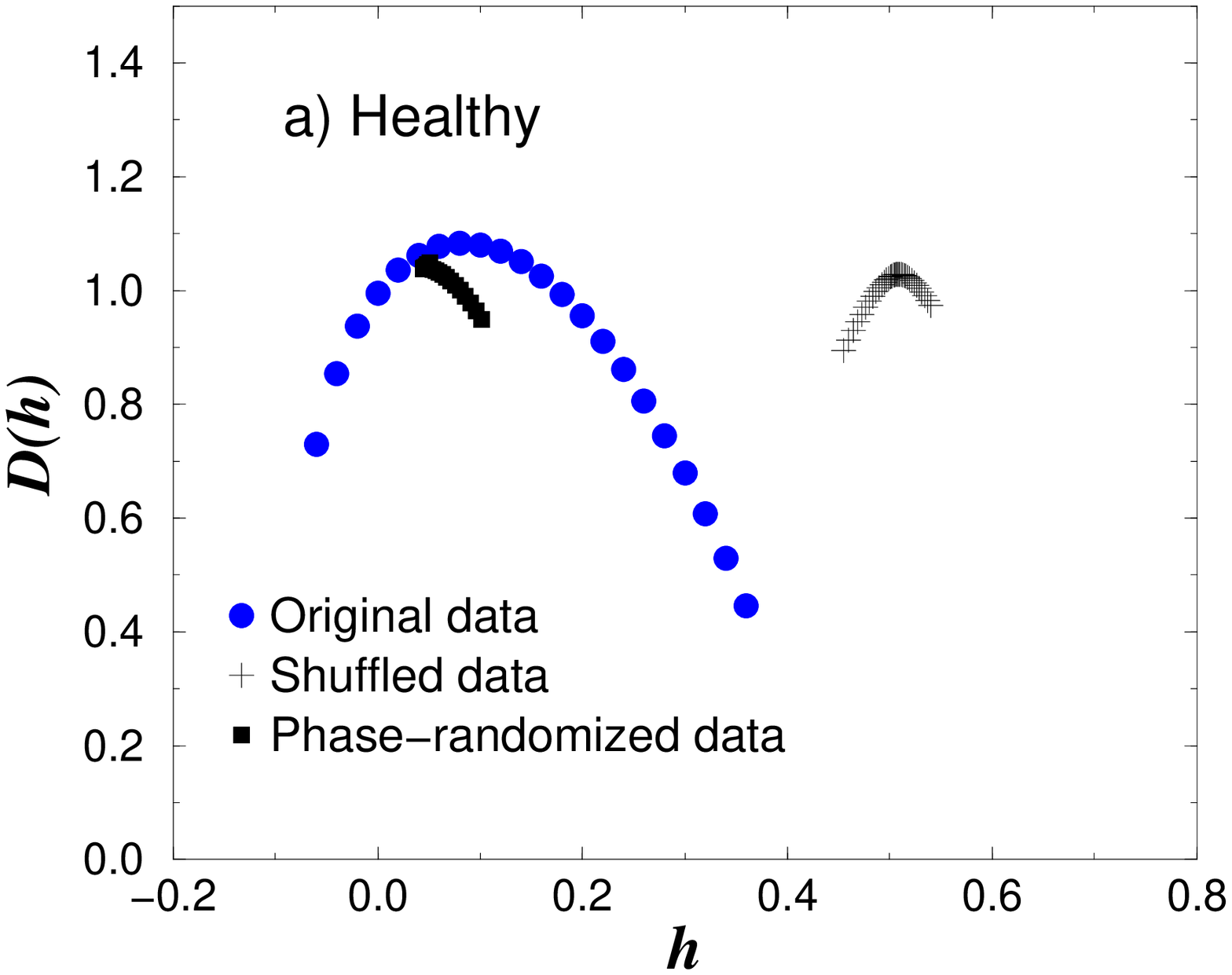}}
\hspace*{-1.0cm}
\epsfysize=0.52\columnwidth{\epsfbox
{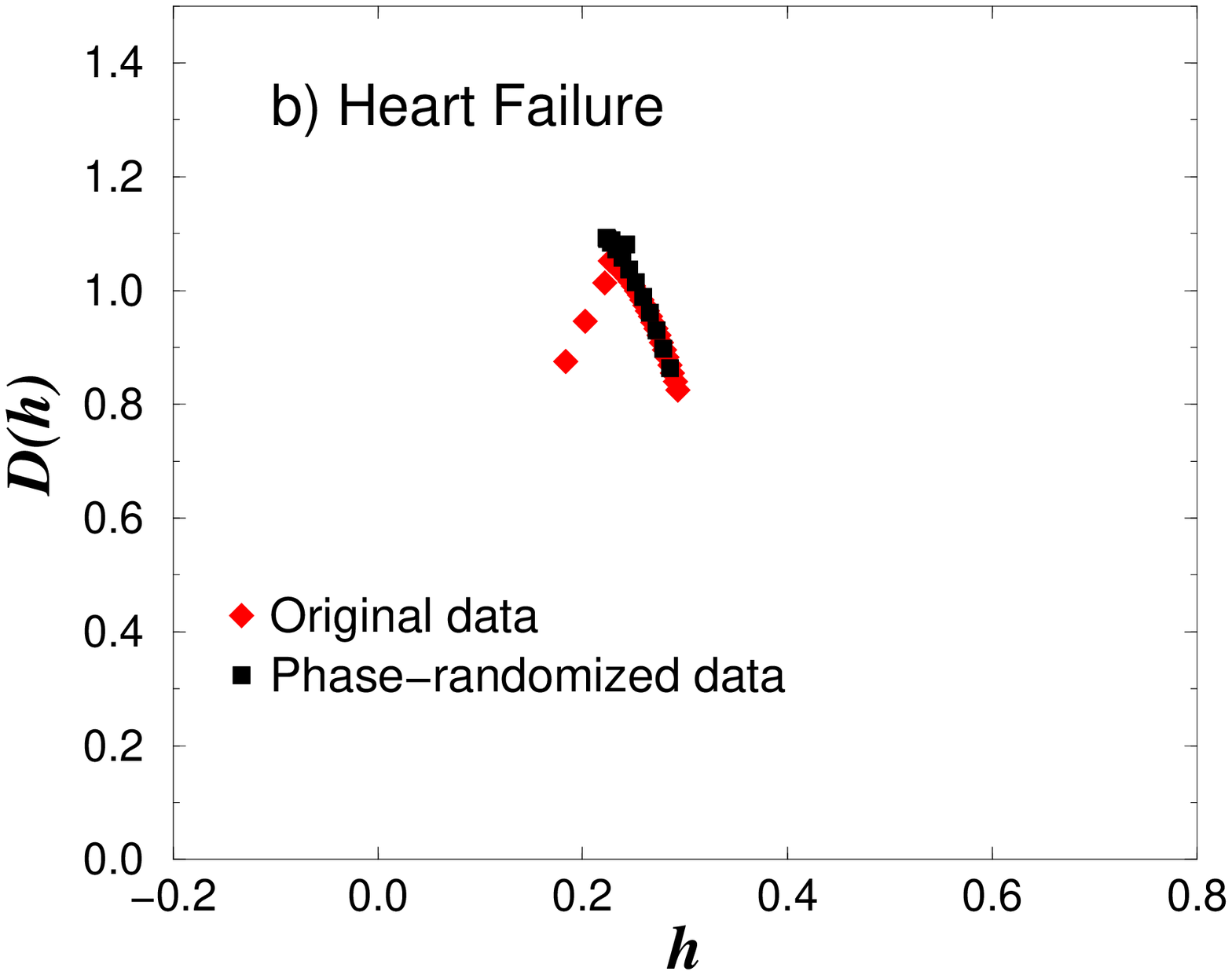}}}
\vspace*{1.0cm}
\caption{ (a) The fractal dimensions $D(h)$ for a 6h daytime record of
a healthy subject. After reshuffling and integrating the increments in
this interbeat interval time series, so that all correlations are lost
but the distribution is preserved, we obtain monofractal behavior ---
a very narrow point-like spectrum centered at {$h \equiv H =
1/2$}. Such behavior corresponds to a simple random walk.  A different
test, in which the $1/f$-scaling of the heart beat signal is preserved
but the Fourier phases are randomized (i.e., nonlinearities are
eliminated) leads again to a monofractal spectrum centered at
$h~\approx~0.07$, since the linear correlations were preserved.  These
tests indicate that the observed multifractality is related to
nonlinear features of the healthy heart beat dynamics rather than to
the ordering or the distribution of the interbeat intervals in the
time series.  (b) The fractal dimensions $D(h)$ for a 6h daytime
record of a heart failure subject. The narrow multifractal spectrum
indicates loss of multifractal complexity and reduction of
nonlinearities with pathology.}
\label{f-test}
\end{figure}

  We systematically compare our method with other widely used methods
of heart rate time series analysis \cite{Peng95,Thurner98,Amaral98}.
Several of these methods do not result in a fully consistent
assignment of healthy versus diseased subjects \cite{CPS-website}.  In
Fig.~\ref{f-discr}a we illustrate the results of our method based on
the multifractal formalism.  Each subject's dataset is characterized
by three quantities: (1) the standard deviation of the interbeat
intervals; (2) the exponent value $\tau(q=3)$ obtained from the
scaling of the third moment $Z_3(a)$, and (3) the degree of
multifractality, defined as the difference between the maximum and
minimum values of local Hurst exponent $h$ for each individual
[Fig.~\ref{f-test}].  We find that the multifractal approach
robustly discriminates the healthy from heart failure subjects.

  We next blindly analyze a separate database containing 10 records, 5
from healthy individuals and 5 from patients with congestive heart
failure.  The time series in the new database are shorter than the
ones in our database; on average they are only 2 hours long ---that
is, less than 8000 beats.  Figure~\ref{f-discr}b shows the
projection on the x-y plane of our data presented in
Fig.~\ref{f-discr}a. Marked in black, we show the results for the
blind test.  Our approach clearly separates the blind test subjects
into two groups: 1, 3, 5, 6, and 10 fall in the healthy group and 2,
4, 7, 8 and 9 in the heart failure group.  Unblinding the test code
reveals that indeed subjects 1, 3, 5, 6 and 10 are healthy, while 2,
4, 7, 8 and 9 are heart failure patients. We conclude that an analysis
incorporating the multifractal method may add diagnostic power to
contemporary analytic methods of heartbeat (and other physiological)
time series analysis.


  The multifractality of heart beat time series also enables us to
quantify the greater complexity of the healthy dynamics compared to
pathological conditions. Power spectrum analysis defines the
complexity of heart beat dynamics through its scale-free behavior,
identifying a single scaling exponent as an index of healthy or
pathologic behavior.  Hence, the power spectrum is not able to
quantify the greater level of complexity of the healthy dynamics,
reflected in the heterogeneity of the signal.  On the other hand, the
multifractal analysis reveals this new level of complexity by the 
broad range of exponents necessary to characterize the healthy
dynamics. Moreover, the change in shape of the $D(h)$ curve for the
heart failure group may provide insights into the alteration of the
cardiac control mechanisms due to this pathology.

  To further study the complexity of the healthy dynamics, we perform
two tests with surrogate time series.  First, we generate a surrogate
time series by shuffling the interbeat interval increments of a record
from a healthy subject.  The new signal preserves the distribution of
interbeat interval increments but destroys the long-range correlations
among them.  Hence, the signal is a simple random walk, which is
characterized by a single Hurst exponent $H = 1/2$ and exhibits
monofractal behavior [Fig.~\ref{f-test}a].  Second, we generate a
surrogate time series by performing a Fourier transform on a record
from a healthy subject, preserving the amplitudes of the Fourier
transform but randomizing the phases, and then performing an inverse
Fourier transform.  This procedure eliminates nonlinearities,
preserving only the linear features of the original time series.  The
new surrogate signal has the same $1/f$ behavior in the power spectrum
as the original heart beat time series; however it exhibits
monofractal behavior [Fig.~\ref{f-test}a].  We repeat this test on a
record of a heart failure subject. In this case, we find a smaller
change in the multifractal spectrum [Fig.~\ref{f-test}b].  The results
suggest that the healthy heartbeat time series contains important
phase correlations cancelled in the surrogate signal by the
randomization of the Fourier phases, and that these correlations are
weaker in heart failure subjects.  Furthermore, the tests indicate
that the observed multifractality is related to nonlinear features of
the healthy heartbeat dynamics. A number of recent studies have tested
for nonlinear and deterministic properties in recordings of interbeat
intervals \cite{Lefebvre,Yamamoto,Kanters,Sugihara,Poon97}. Our
results are the first to demonstrate an explicit relation between the
nonlinear features (represented by the Fourier phase interactions) and
the multifractality of healthy cardiac dynamics [Fig.~\ref{f-test}].

  From a physiological perspective, the detection of robust
multifractal scaling in the heart rate dynamics is of interest because
our findings raise the intriguing possibility that the control
mechanisms regulating the heartbeat interact as part of a coupled
cascade of feedback loops in a system operating far from equilibrium
\cite{Sreenivasan87,Ivanov98}.  Furthermore, the present results
indicate that the healthy heartbeat is even more complex than
previously suspected, posing a challenge to ongoing efforts to develop
realistic models of the control of heart rate and other processes
under neuroautonomic regulation.

\vfill
\eject
\centerline{\bf Methods}

  Heart failure (CHF) data were recorded using a Del Mar Avionics Model
445 Holter recorder and digitized at 250 Hz. Beats were labeled
using ``Aristotle'' arrhythmia analysis software. By using criteria
based on timing and QRS morphology, ``Aristotle'' labels each detected beat
as normal, ventricular ectopic, supraventricular ectopic, or unknown
and the location of the R-wave peaks is determined with a resolution
of 4 ms.

  Healthy datasets were recorded using a Marquette Electronics Series
8500 Holter Recorder. Using a Marquette Electronics Model 8000T Holter
Scanner, the tapes were then digitized at 128 Hz., scanned and
annotated. The annotations were manually verified by an experienced
Holter scanning technician. The location of the R-wave peaks was thus
determined to a resolution of 8 ms.

  The finite resolution implies that our estimates of the interbeat
intervals are affected by a white noise due to estimation error.  The
signal-to-noise ratio is for both healthy and heart failure of the
order of 100.  Furthermore, the white noise due to the measurements
would lead to the detection of a local Hurst exponent $h=0$ at very
small scales.  For that reason, in this study we have considered only
scales larger than 16 beats.

  From the beat annotation file, only the intervals (NN) between
consecutive normal beats were determined; thus intervals containing
non-normal beats were eliminated from the NN interval series. For the
heart failure (CHF) data an average of 2\% (0.1\% min to 0.6\% max) of
the intervals were eliminated, and for the normal data an average of
0.01\% (0\% min to 0.06\% max) were eliminated. No interpolation was
done for eliminated intervals.

  In order to eliminate outliers due to missed QRS detections which
would give rise to erroneously large intervals that may have been
included in the NN interval series, a moving window average filter was
applied. For each set of 5 contiguous NN intervals, a local mean was
computed, excluding the central interval. If the value of the central
interval was greater than twice the local average, it was considered to
be an outlier and excluded from the NN interval series. This criterion
was applied to each NN interval in the series. For the CHF data an
average of 0.02\% (0\% min to 0.1\% max) of the intervals were
eliminated and for the normal data an average of 0.07\% (0\% min to
0.7\% max) were eliminated. No interpolation was done for eliminated
intervals.
Overall a total of 2\% (0.1\% min to 6\% max) of the total number of
RR intervals were eliminated for the CHF data and 0.08\% (0\% min to
0.7\% max) for the normal data. 

  Next, we build a time series of increments between consecutive NN
intervals and calculated their standard deviation.  We then identified
all pairs of associated increments with opposite signs and with an
amplitude larger than 3 standard deviations.  The values of the
increments for each pair were replaced by linear interpolations of their
values and the time serie of NN interval time was reconstructed by
integration from the filtered increments. About 1\% of time intervals
were corrected by this procedure.



\vfill
\eject

\vfill
\eject

\vfill
\eject

\end{document}